\documentclass[namedreferences]{SolarPhysics}
\usepackage[optionalrh]{spr-sola-addons} 
\usepackage{graphicx}        
\usepackage{color}           
\usepackage{url}             

\begin{document}

\begin{article}

\begin{opening}

\title{STUDY OF CME PROPERTIES USING HIGH RESOLUTION DATA}

\author{V.G.~\surname{Fainshtein}$^{1}$\sep
        Ya.I.~\surname{Egorov}$^{2}$\sep
       }
\runningauthor{Fainshtein V.G, Egorov Ya.I.}
\runningtitle{STUDY OF CME PROPERTIES USING HIGH RESOLUTION DATA}

   \institute{$^{1}$ ISTP SB RAS
                     email: \url{vfain@iszf.irk.ru} \\
              $^{2}$ ISTP SB RAS
                     email: \url{egorov@iszf.irk.ru} \\
             }

\begin{abstract}
The joint use of high-resolution data from PROBA2 and SDO satellites and LASCO/SOHO coronographs enabled us to examine early stages of initiation and propagation of six limb CMEs registered in June 2010 - June 2011. For five events under consideration, the CME initiation is marked by filament (prominence) eruption or by a loop-like structure having another nature.  Subsequently, several loop-like structures having higher brightness and following each other at different velocities appear in the region of the CME initiation. The CME frontal structure is formed by these loop-like structures. The time-dependent velocities and acceleration of the ejection front have been obtained for all CMEs under consideration. We have drawn a conclusion about the possible existence of two CME types dependent on the time profile of their velocity. The first CME type comprises the ejections whose velocity decreases abruptly by more than 100 km/s after having reached the maximum; it thereupon passes to slow deceleration. The second CME type is formed by the ejections whose velocity varies insignificantly after reaching the maximum. The CME angular size is shown to increase up to threefold at the initial stage of propagation; it increases twofold 3.5-11 minutes after the first measurement of this parameter. When considering 3 CMEs, we see that their broadening exceeds their extension in the longitudinal direction during a certain period of time at the initial propagation stage.
\end{abstract}
\end{opening}

\section{Introduction}
     \label{S-Introduction}

	Physical mechanisms of generation of coronal mass ejections (CMEs) still remain unclear in many respects \cite{Forbes}. This burning issue can be addressed due to the use of data with high temporal and spatial resolution that will allow us to study initiation and initial propagation stage of CMEs. In 2010, Solar Dynamics Observatory (SDO) \cite{Pesnell} was launched with the on-board Atmospheric Imaging Assembly (AIA) instrument \cite{Lemen} having unique characteristics. AIA enables us to observe the Sun in the wavelength range from extreme ultraviolet ($\lambda=9.4 \; nm$ ) to ultraviolet ($\lambda=170.0 \; nm$ ) with high temporal (10 seconds) and spatial resolutions (0.6 arcsec pixels). Consequently, it will be possible to reveal regularities of the CME generation and the initial phase of the CME propagation. 
	
The additional use of data from the PROBA2 (Project for On-Board Autonomy) satellite launched in 2009 will contribute to this study, too (\url{http://proba2.oma.be/index.html/about/}). The on-board instrument SWAP \\
\cite{Katsiyanni} observes the Sun in extreme ultraviolet at 17.4 \textit{nm}, whereas LYRA (\url{http://proba2.oma.be/index.html/science/}) monitors the solar irradiance in the wavelength range from soft X-rays to ultraviolet. Though the temporal (1 minute) and spatial (3.17{''}) resolutions of SWAP are worse than those of AIA instruments, it has a big advantage: its field of view is significantly larger (54 arcmin as against 41 arcmin (SDO)). 

	Establishment of experimental regularities of the CME propagation immediately after its initiation is of importance when studying CME features. Over the last decade, several studies were devoted to different aspects of the initial stage of the CME propagation, using data from TRACE, SOHO, STEREO, GOES satellites (see papers \cite{Gallagher, Zhang, Maricic09}). First of all, kinematics of the CME propagation was examined. Three periods of the initial propagation of CMEs were established \cite{Zhang}: slow increase in velocity within several tens of minutes, rapid increase in velocity (the main CME acceleration) within the period from several minutes to several tens of minutes (sometimes up to several hours) when the acceleration amplitude reaches the value of several $km/s^2$ in some ejections. The third period is the propagation phase of CMEs (i.e., the 'quiet' propagation of CMEs with an acceleration that is small or moderate in absolute magnitude).
	 
	A close connection of the CME acceleration with energy release was established in the CME-related solar flares \cite{Zhang, Maricic07, Temmer08, Temmer10}. There was a positive correlation between the duration of the main CME acceleration tACC and the time of increase tSXR in intensity of soft X-rays ISXR from the CME-related flare area up to maximum values \cite{Zhang, Maricic07}. The CME acceleration $a(t)$ and hard X-ray emission $I_{HX}(t)$ (or its related derivative $dI_{SXR}/dt$) from the flare area were shown to be often synchronised \cite{Maricic07, Temmer08, Temmer10}. An inverse correlation dependence of acceleration $V_{max}/t_{acc}$ on $t_{acc}$ (or $V_{max}/t_{SXR}$ and $t_{SXR}$) was established \cite{Zhang, Vrsnak}. 

	Due to unique capabilities of the instruments on board SDO, new results on the CME initiation and its initial propagation stage have been obtained \cite{Patsourakos2}. With a CME as an example, we have shown that the CME initiation has three stages: the stage of slow self-similar expansion, short-term strong hyperexpansion in the transverse direction, and one more stage of self-similar expansion in the SDO field of view. There is also a strong short-term CME acceleration with a subsequent deceleration. Note that the conclusion about the existence of a period of the CME hyperexpansion in the transverse direction was drawn from PROBA2 data later in \cite{Fain}.  
	
	In this paper, we go on studying the CME initiation during the new solar cycle, using SDO/AIA data. We also examine the initial stage of the CME propagation with the use of data from the SDO, PROBA2 and SOHO spacecraft. The goal of this study is to get new experimental results concerning the CME kinematics and time variations of their geometric parameters that could be used to construct adequate models of the CME generation and propagation at the initial stage.

\section{Data and investigation techniques} 
      \label{Data and investigation techniques}      

	We have analysed CMEs of the new solar cycle, recorded during the period from June 2010 to June 2011. The events to analyse had to meet the following requirements: 
\begin{enumerate}
	\item two spacecraft - SDO and PROBA2 - must observe CMEs. Resolution of the corresponding CME images to analyse must be sufficiently high. In particular, at least 4 consecutive images must represent each event in order to enable position of the CME front be determined and time-dependent velocity and CME acceleration to be calculated;
	\item events must be related to X-ray flares; 
	\item events must be limb or close to limb ones;
	\item each ejection selected using SDO and PROBA2 data must correspond to a CME recorded in the LASCO field of view, according to the catalogue data .
\end{enumerate} 

	There were 9 events that met these requirements. CME observed on 18 August 2010 was not considered, since its initiation and propagation were complex. We did not consider the non-limb event of 14 August 2010 and the event of 24 February 2011 either, since interpretation of its initial data still goes on. An event was considered to be limb, if the CME-related flare was at a distance $\geq 60^{\circ}$ from the centre of the solar disc (in the heliographic coordinate system). There were 5 such events: on 13 May 2010, 11 February 2011, 08 March 2011, 27 April 2011, and 07 June 2011. The flare position in 2011 was determined as the position of the source of X-ray emission in the range from 6 to 12 keV, according to Reuven Ramaty High-Energy Solar Spectroscopic Imager (RHESSI) data (\url{http://sprg.ssl.berkeley.edu/~tohban/browser/}).  The event of 27 March 2011 was difficult to relate to a flare. Its source was probably just behind the limb. We estimated latitude of this CME's source: it was $\approx$ $25^{\circ} - 35^{\circ}$ relative to the limb. 
	
	The CME initiation and propagation at the initial stage have been studied using SDO/AIA data and FITS images of the Sun (\url{http://jsoc.stanford.edu/ajax/lookdata.html}), data processing level 1 in EUV lines $\lambda = 17.1\;nm$ and 21.1 \textit{nm} that are formed when plasma temperature is $\approx$ 9 105 K and 2 106 K. To analyse the CME kinematics, we used calibrated solar images (\url{http://proba2.oma.be/swap/data/bsd}) taken by SWAP in the emission line having a wavelength of 17.4 \textit{nm}.  
To determine CMEs using SDO/AIA data, difference images were plotted for each pixel with the use of the following relation: 
$$
\Delta I = (I_2 \frac{T_{exp0}}{T_{exp2}}-I_1 \frac{T_{exp0}}{T_{exp1}})/I_0
$$
Here, $I_2$ is the image brightness at a later moment of time; $I_1$, at one of the previous moments of time; $I_0$, at the third moment of time which is earlier than $I_1$. We took account of the fact that exposure time $T_{exp1}$, $T_{exp2}$ and $T_{exp0}$ can be different when obtaining three images. To determine CMEs from SWAP data, we used only difference $ \Delta I = I_2-I_1$.

	Data (\url{http://sharpp.nrl.navy.mil/cgi-bin/swdbi/lasco/images/form}) from Large Angle and Spectrometric Coronagraphs (LASCO) aboard SOHO were used to examine the CME propagation at distances $R>2R_0$ ($R$ is counted from the centre of the solar disc in the plane of the sky, $R_0$ is the solar radius). To get information about soft X-ray emission, we used data from Geostationary Operational Environmental Satellites (GOES) (\url{http://goes.ngdc.noaa.gov/data}) and PROBA2/LYRA (\url{http://proba2.oma.be/index.html/science}). For one of the events (on 08 March 2011), we compared the time variation in the main acceleration with the time dependence of hard X-rays recorded by RHESSI (\url{http://sprg.ssl.berkeley.edu/~tohban/browser/}) over the energy range from 25 to 50 KeV. 
	
	Information about eruptive filaments (prominence) from the closest proximity to the CME generation region was provided by the telescopes observing the Sun in the H-alpha line (\url{http://swrl.njit.edu/ghn_web/}) and by Coronado SolarMax H-alpha Imager at Mauna Loa Solar Observatory (MLSO) (\url{http://mlso.hao.ucar.edu/mlso_data_CORONADO_2010.html}).  
	
	The CME velocity $V_i$ was calculated using formula 
	$$
		V_i=\frac{L^F_{i+1}-L^F_i}{t_{i+1}-t_i}
	$$
	where $L^F_i$ is the distance from the CME front to the chosen point on the visible solar disc, generally along the curved trajectory, at moment of time $t_i$. We supposed that velocity $V_i$ was at the moment $t^F_{i+1} = (t_{i+1} + t_i)/2$. Obtained values $V_i(t)$ were approximated by B-splines, taking account of the error in determining $V_i(t)$ at each moment of time. This resulted in the smooth function $V(t)$. In the field of view of LASCO coronagraphs, we determined the HCME front position from the catalogue (\url{http://cdaw.gsfc.nasa.gov/CME_list}). The CME acceleration was determined from ratio $a(t)=dV(t)/dt$. The minimum error in calculating velocity was determined from ratio $ \Delta V_{min} = 2 \delta P/(t_{i+1}-t_i)$, where $\delta P$ is the spatial resolution of the instrument. The CME front position was determined visually from solar images, whereas the error in calculating velocity was determined as root-mean-square deviation of the calculated CME-front velocity from the mean value during N measurements (N = 5 - 10).  We supposed that the CME front position could be determined accurate within $0.5R_0$ in the field of view of LASCO at $R >5R_0$ \cite{Shanmugaraju}.    
	
	To determine CME velocity $V(t)$, we used not only values of the CME front velocity obtained using data from a certain instrument, but also 'intermediate' points $V(T_{1,2}) = (R(T_2)-R(T_1))/(t_{T_2}-t_{T_1})$, where $R(T_2)$ is the CME front position at moment $t_{T_2}$ of its first registration by telescope $Т2$, $R(T_1)$ is the CME front position at moment $t_{T_1}$ of its last registration by telescope $Т1$. For instance, $T_1$ - SDO/AIA, $T_2$ - LASCO C2.
	
	To determine reliably the eruptive prominence (flux ropes, according to our terminology), loop-like structures that form CMEs, and the CME frontal structure, special films were made, which allowed us to determine the structures that moved steadily. In what follows, we will use the term 'flux rope' to denote loop-like structures from hot coronal matter.  To determine the magnetoplasma flux rope that preceded the eruption onset several tens of minutes and that would later form a CME, we used source images of the Sun after special processing. Numerous coronal loops were observed at these images. The relevant loop was at the place where it would be later observed as an eruptive flux rope on difference images.

\section{Results} 
      \label{Results}      

 \begin{figure}    
   \centerline{\includegraphics[width=120mm]{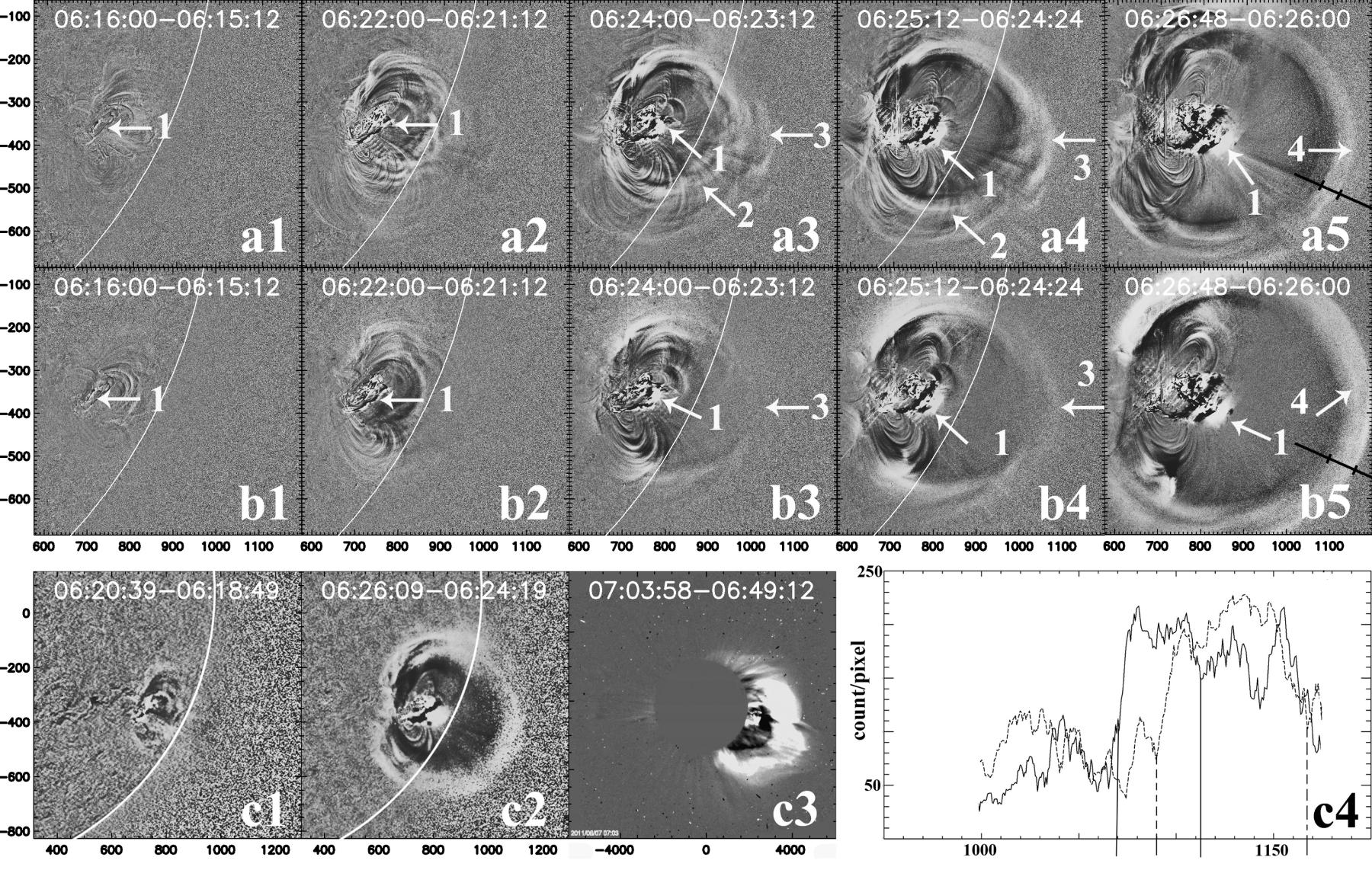}}
              \caption{The event of 07 June 2011. a1 - a5 are the difference solar images obtained using SDO/AIA data in the spectral line $\lambda=17.1\; nm$; b1 - b5, in the lines $\lambda=21.1\; nm$; c1 - c2 are the difference solar images obtained using PROBA2/SWAP data in the line $\lambda=17.4\; nm$; c3 is the difference image of the solar corona, according to LASCO C2 data; c4 are the brightness scans along segments of the straight black line presented in Fig. a5 and b5. Solid line in Fig. c4 denotes the brightness scan in Fig. a5; dashed line, in Fig. a6. Arrows No. 1 denote the eruptive prominence; arrows No. 2, the inner loop when CME is under formation; arrow No. 3, the outer loop; arrow No. 4, the formed frontal structure. Each image is provided with a coordinate axis allowing us to determine location of any element on the image, relative to the centre of the solar disc, in terms of seconds of arc. In Fig. a5 and b5, dashes perpendicular to segments, along which brightness was scanned, denote 
inner and outer boundaries of the region of the most significant variation in brightness in the CME frontal structure.  In Fig. c4, they are marked off by vertical lines: solid ones for the image in the line $\lambda=17.4\; nm$; dashed ones, in the line $\lambda=21.1\; nm$.}
   \label{fig1}
   \end{figure}
   
     \begin{figure}    
   \centerline{\includegraphics[width=120mm]{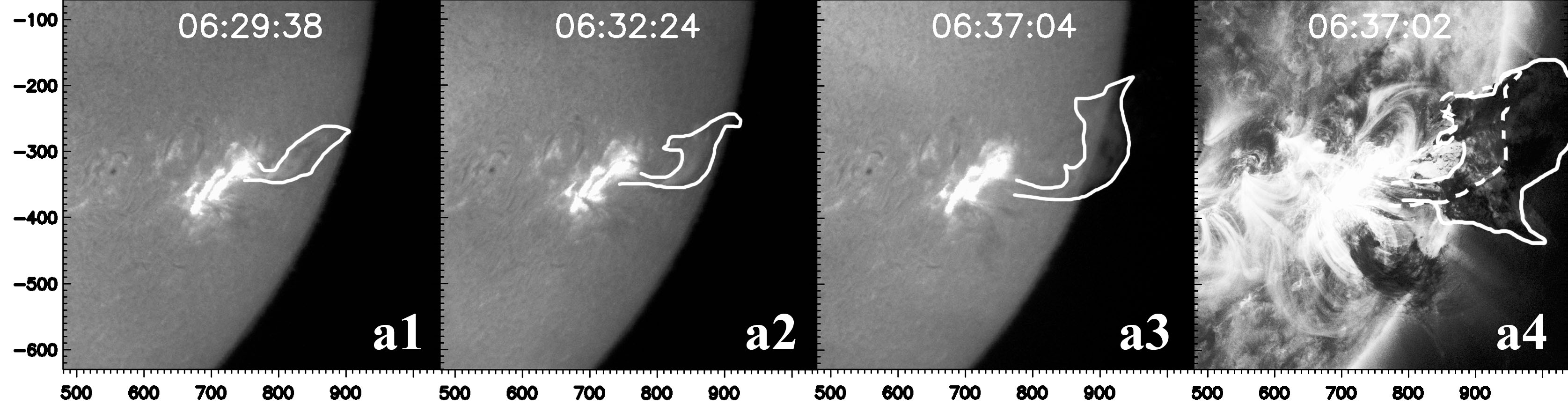}}
              \caption{(a1 - a3) is the eruption of filament related to CME of 07 June 2011, according to measurements in the H-alpha line. The white line indicates the boundary of the eruptive filament. a4 is the comparison between eruptive filaments in the H-alpha line (white dashed line) and in the line $\lambda=17.1\; nm$ (white solid line) at 06:37:02 (for the line at 17.1 \textit{nm}) and at 06:37:04 (for the H-alpha line). Coordinate values on the coordinate axes are in terms of seconds of arc from the centre of the solar disc.}
   \label{fig2}
   \end{figure}
   
   \begin{figure}    
   \centerline{\includegraphics[width=120mm]{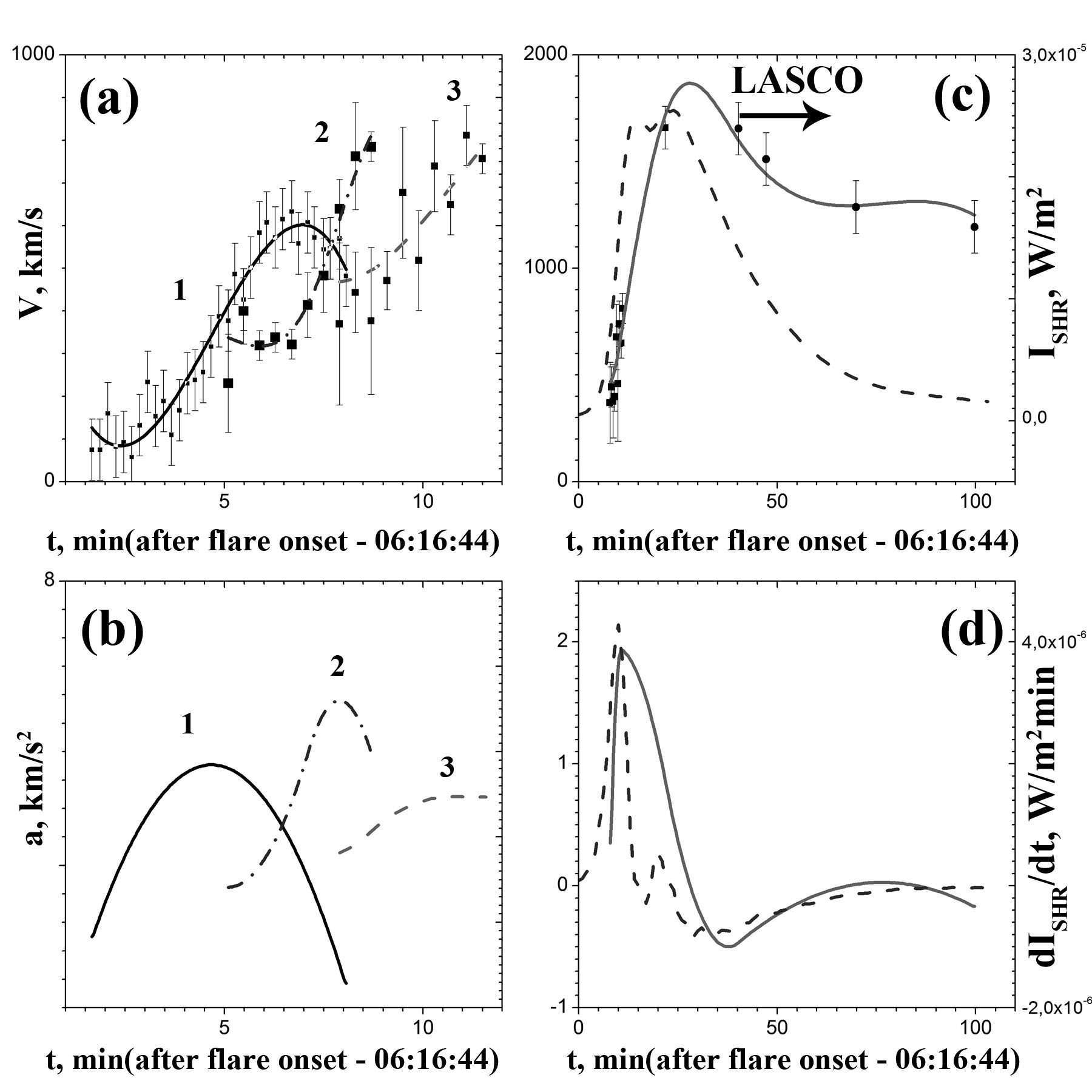}}
              \caption{(a) is the time dependence of velocity of the eruptive prominence (solid line) and of two loop-like CME structures (dotted line denotes the outer structure; dash-and-dot line, the inner one), according to SDO/AIA data for the event of 07 June 2011; (b) is the time dependence of acceleration of the said structures; (c) is velocity V(t) obtained from joint SDO/AIA (squares), PROBA2/SWAP (triangles) and SOHO/LASCO (circles) data - solid line, and time-dependent soft X-ray intensity ISXR(t) - dashed line;  (d) is the acceleration of the CME frontal structure a(t) - solid line - and time dependence dISXR/dt - dashed line.}
   \label{fig3}
   \end{figure}
   
   \begin{figure}    
   \centerline{\includegraphics[width=120mm]{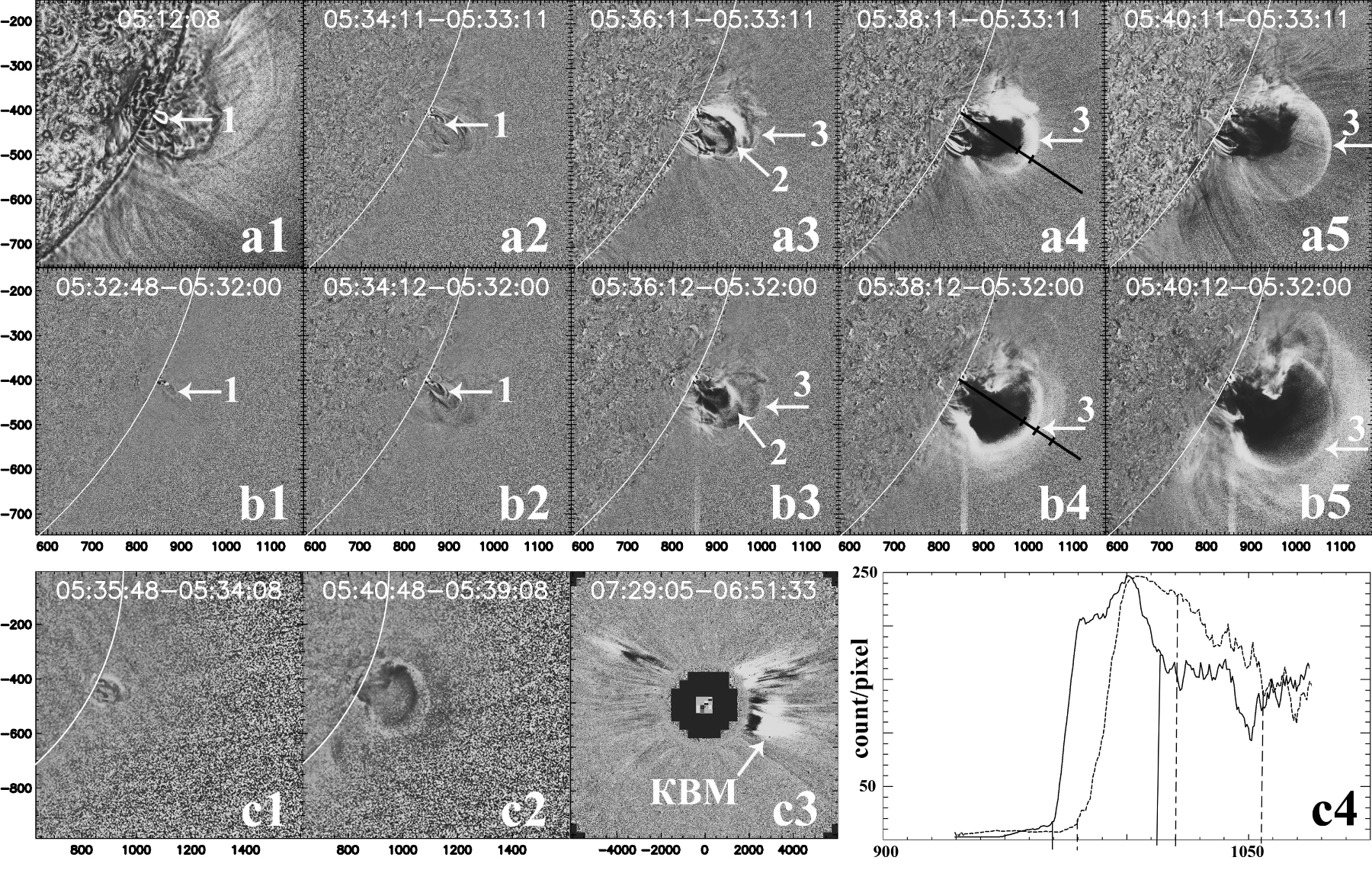}}
              \caption{The same as in Fig. 1 but for CME on 13 June 2010. Unlike CME of 07 June 2011, initiation of CME in this case is related to eruption of flux rope whose nature is different from that of the filament (prominence). Label "1" on a1, b1, b2 is the flux rope (outlined) before initiation of the CME-related flare, registered in the line $\lambda=19.3\; nm$.}
   \label{fig4}
   \end{figure}
   
   \begin{figure}    
   \centerline{\includegraphics[width=120mm]{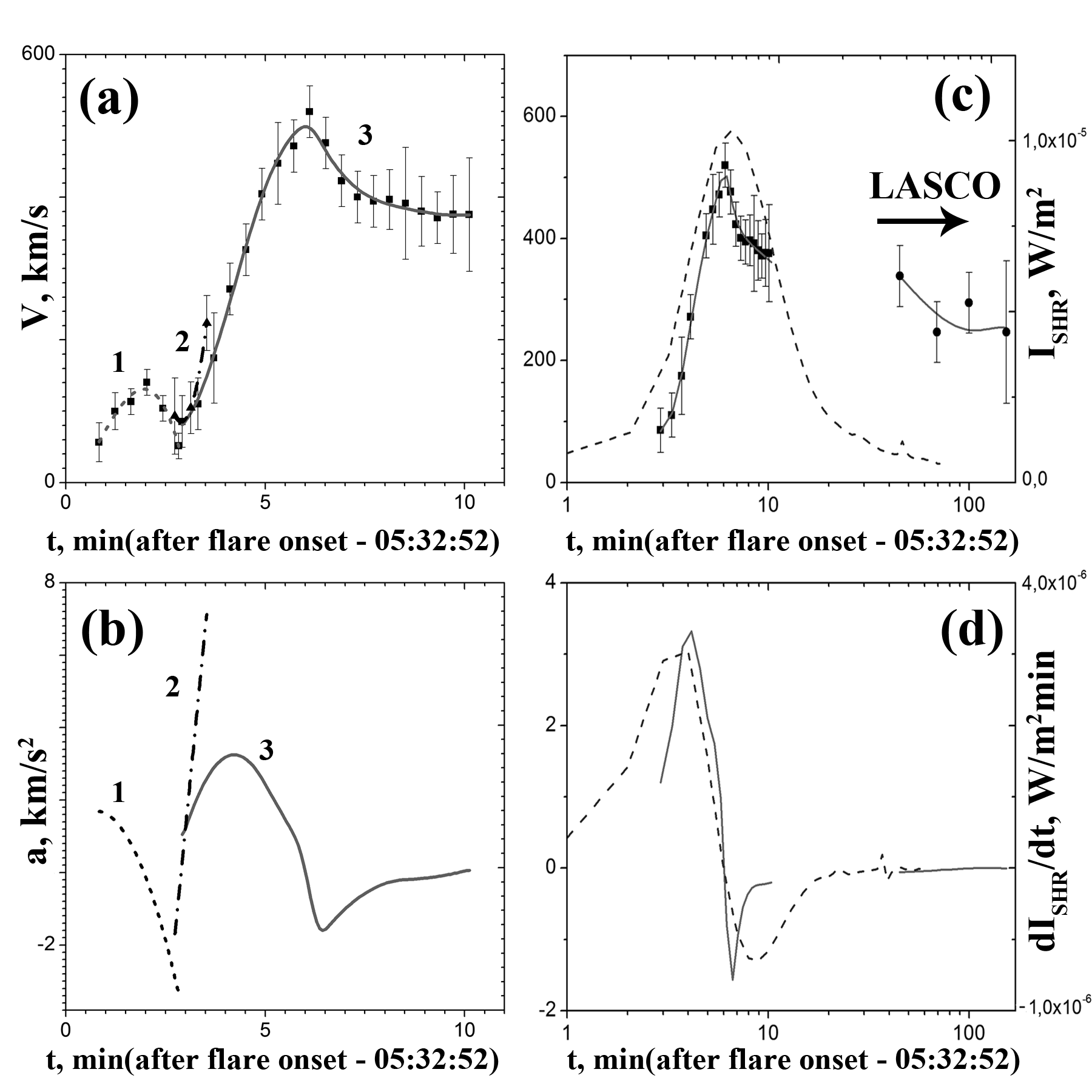}}
              \caption{The same as in Fig. 3 but for CME on 13 June 2010. Solid line in (a - b) denotes time dependence of velocity and of acceleration of the outer CME loop; dashed line, velocity and acceleration of the eruptive flux rope; dash-and-dot line, velocity and acceleration of the inner loop, according to SDO/AIA data. }
   \label{fig5}
   \end{figure}
   
   \begin{figure}    
   \centerline{\includegraphics[width=120mm]{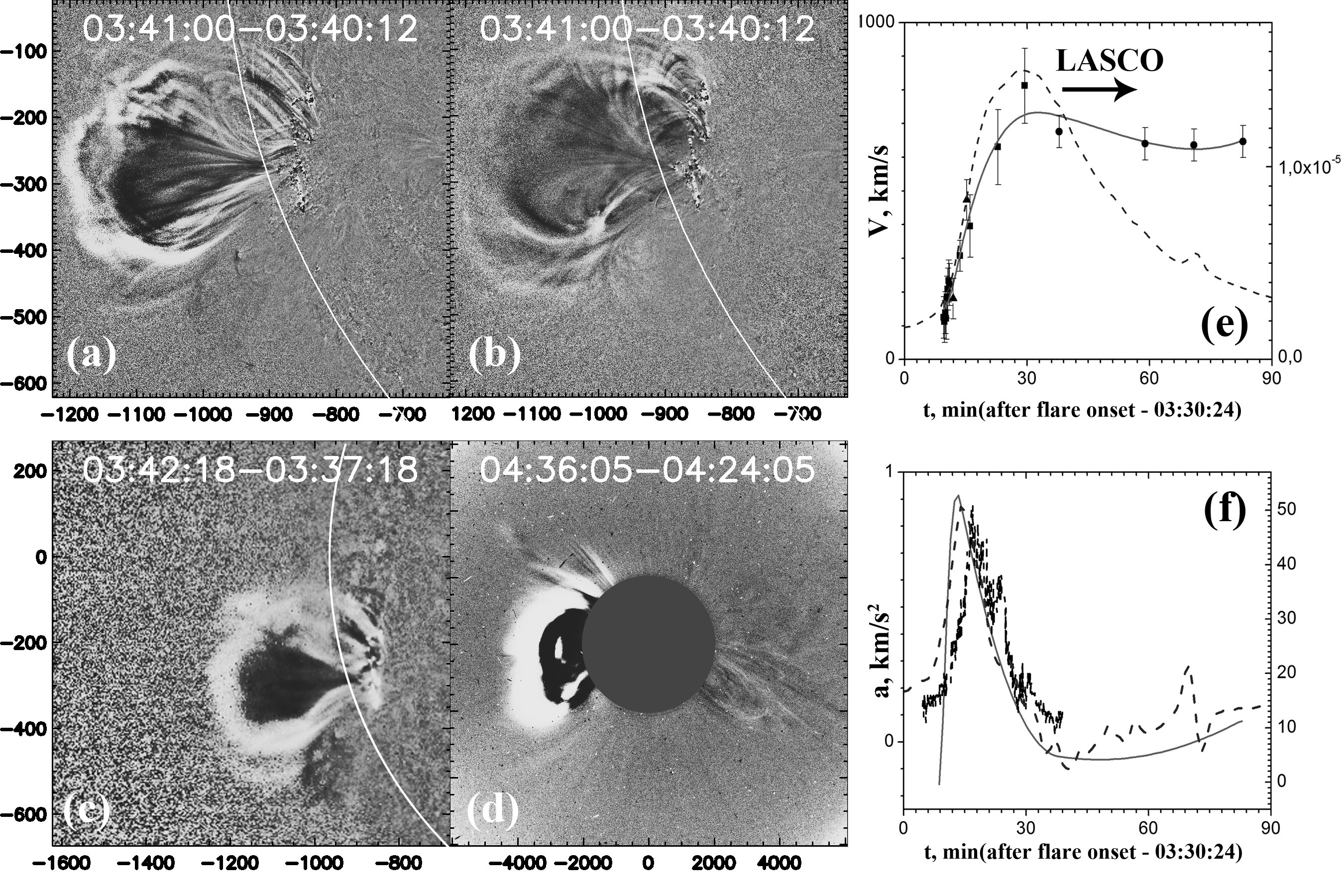}}
              \caption{Difference images of CME of 08 March 2011, when CME is already formed, according to SDO data in the lines at 17.1 nm (a) and 21.1 nm (b); according to PROBA2 data (c); according to LASCO C2 data (d). (e) is the time dependence of the CME front velocity, according to joint data from SDO/AIA, PROBA2/SWAP and SOHO/LASCO (continuous dependence of velocity on time is denoted by solid line), and the time profile of soft X-ray intensity ISXR(t) - dashed line; (f) are the CME acceleration profiles (solid line) and dISXR/d t - dashed line. Vertical dashes show the time dependence of hard X-ray emission.}
   \label{fig6}
   \end{figure}
   
   \begin{figure}    
   \centerline{\includegraphics[width=120mm]{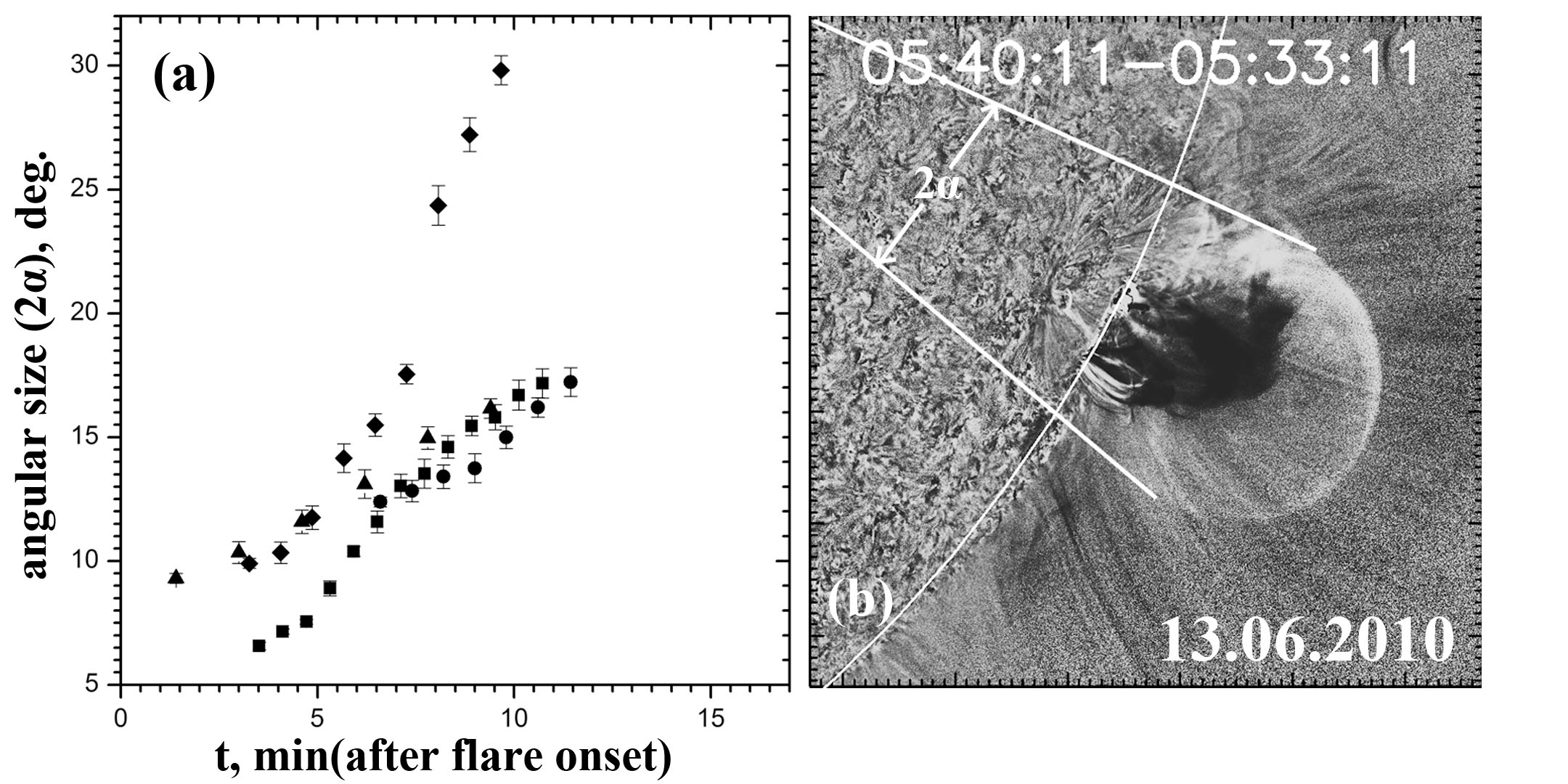}}
              \caption{ (a) is the time dependence of the CME angular size $2 \alpha$ for four CMEs. ; (b) illustrates determination of the CME angular size $2 \alpha$. Vertex of the  angle determining the CME size is placed in the centre of the solar disc. }
   \label{fig7}
   \end{figure}
   
   \begin{figure}    
   \centerline{\includegraphics[width=120mm]{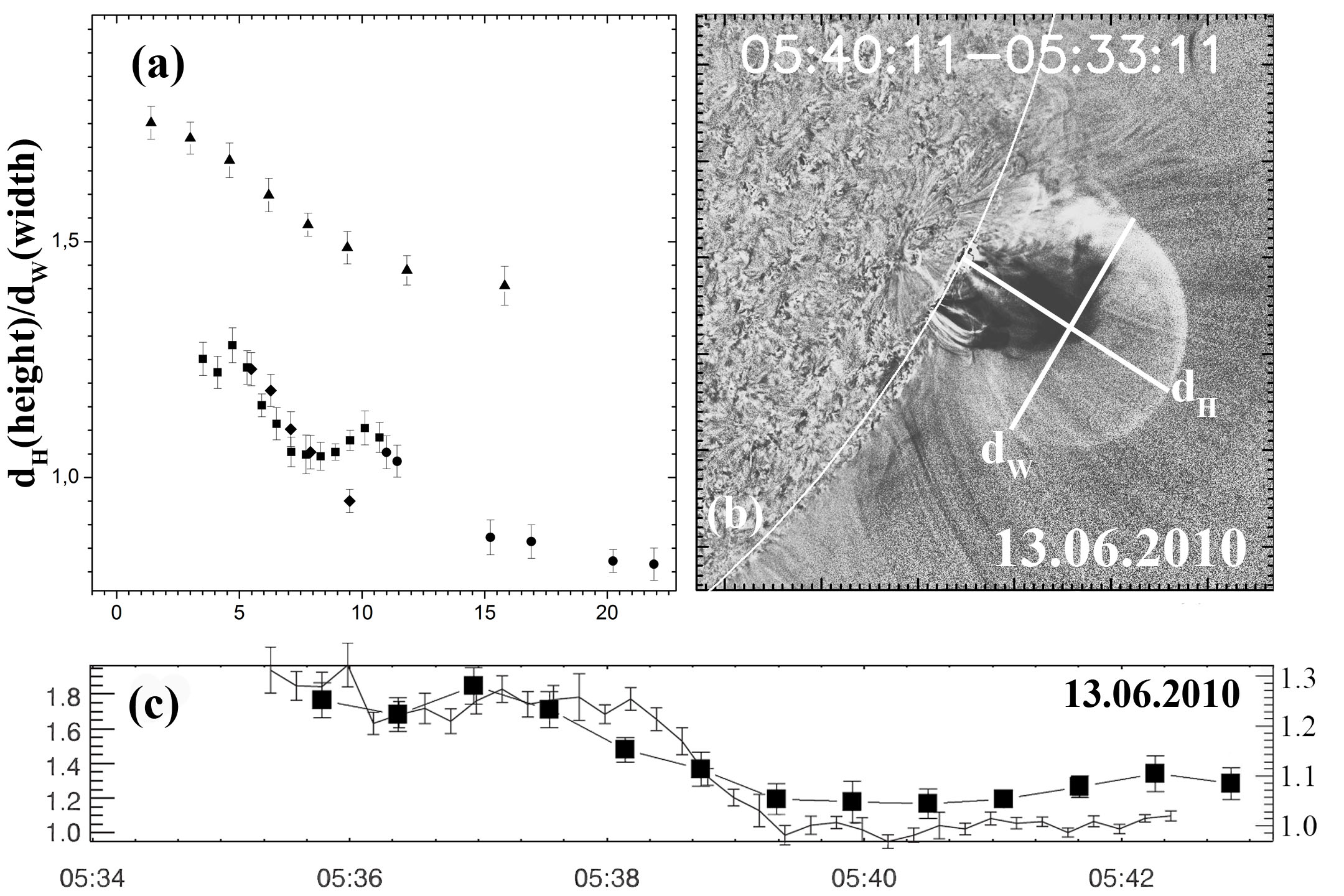}}
              \caption{ (a) is the time variation in ratio of the CME height $d_H$ to its width $d_W$ (square - 13 June 2010; circle - 08 March 2011; triangle - 27 April 2011; diamond - 07 June 2011); (b) demonstrates the CME height and width; (c) is the comparison between $[d_H/d_W](t)$ (solid squares) and time dependence of 'aspect ratio'. Scale on the vertical axis on the left in (c) represents 'aspect ratio'; scale on the right, $d_H/d_W$.}
   \label{fig8}
   \end{figure}   

Fig. 1 exemplifies CMEs on 07 June 2011, according to observations of solar radiation in wavelengths of 17.1 and 21.1 nm (SDO/AIA data), and CMEs recorded by SWAP/PROBA2 and LASCO C2 coronagraph. Analysis of all solar images available from SDO/AIA instruments has revealed that initiation of this CME is due to motion of a small structure (see arrows №1 in Fig. 1). As we shall see later, this structure is an eruptive filament (prominence). When CME is under formation, the eruptive filament is preceded by numerous motionless and moving loop-like structures that show an increased brightness. In two lines, sets of such loops, however, are observed in a different way, and they vary over time in a different way. The moving loops have different velocities, and, as we shall see subsequently, some inner loops move at higher velocities than the outer ones. The inner loops finally catch up with the outer ones, thus leading to formation of a frontal structure (i.e., the outer region of the formed CME, showing an increased brightness). Let us compare the CME frontal structure in the two lines at the moment of time $t=$06:26:48. In the line at 17.1 \textit{nm}, the frontal structure is closer to the ejection source than that in the line at 21.1 \textit{nm}. The brightest part of the frontal structure is in the line at 17.1 nm (see Fig.1 (c4)). 

CME of 07 June 2011 is related to the filament eruption, which is confirmed by Fig. 2. It illustrates the filament (prominence) eruption from the region of the CME initiation, according to H-alpha line observations (data from Kanzelhohe Solar Observatory \url{http://www.kso.ac.at/index_en.php}) (there are no H-alpha data during the onset of the filament eruption). Fig. 2(d) shows the eruptive prominence observed in the H-alpha line and superposed onto the loop recorded in the line at 17.1 \textit{nm}. Similarity in location of structures in different lines confirms that the moving line observed in extreme ultraviolet is the eruptive filament.  

Fig. 3(a) presents time dependence of velocity of the eruptive filament (EF) and of two loop-like CME structures that occurred after initiation of the EF motion. Fig. 3(b) demonstrates time profiles of their acceleration. These loop-like structures (inner and outer ones) are marked off by arrows 2 and 3 in Fig. 1. Results obtained using SDO data in the line $\lambda=17.1\; nm$ are displayed in Fig. 3(b). Apparently EF started moving before the flare, since its velocity was $ \geq 100 km/s$ one minute after the flare initiation. One can see that velocity of the earlier loop increased with time and exceeded that of the later loop; the first loop came up with the second one, thus leading to formation of the CME frontal structure. Note that beginning of the outer loop's motion was measured after beginning of the inner loop's motion. Till this moment, the outer loop was observed as a structure showing a weak brightness (see Fig. 1(a2, b2)); its velocity was hard to determine. In what follows, the moment when we first determined translational velocity of the leading edge of its frontal structure (the CME front) will be referred to as the initial moment of propagation of the CME front.  

Fig. 3(c) shows time dependences of the CME velocity V(t) obtained using joint data from AIA, PROBA2 and LASCO, and soft X-ray intensity ISXR(t) from the CME-related flare region. Time profiles $a(t) =dV(t)/dt$ and \\
$[dI_{SXR}/dt](t)$ are presented in Fig. 3(d). Comparison between $V(t)$ and $I_{SXR}(t)$ profiles reveals that the time of increase in the CME velocity up to maximum value is comparable to the time of soft X-ray intensity enhancement (from the moment of the flare initiation to the moment of $I_{SXR}$ maximum); time of $V(t)$ maximum is close to that of $I_{SXR}(t)$ maximum. These results agree with those in \cite{Zhang, Maricic07}. Fig. 3(d) proves that time of $a(t)$ maximum is close to that of $[dI_{SXR}/dt](t)$ maximum. This result agrees with that in \cite{Maricic07}.
 
In \cite{Temmer08, Temmer10}, time profile of the CME acceleration was compared to time variation in hard X-ray intensity $I_{HX}(t)$.  There were no sufficient data on hard X-ray emission from the flare region for that event; therefore, $a(t)$ acceleration was compared to $[dI_{SXR}/dt](t)$. Such a change was possible due to the Neupert effect \cite{Neupert}. According to this effect, time variation in $I_{HX}(t)$ or in intensity of microwave radiation during the impulsive phase is close to $dI_{SXR}/dt$ time profile. 

Noteworthy is the fact that, after having reached its maximum value, the ejection velocity decreases fast ($\approx$ 20 minutes) by several hundreds of $km/s$; it thereupon passes to slow time variation. This influences also the time profile of the CME acceleration, see Fig. 3(d). 
Fig. 4 and 5 present CME recorded on 13 June 2010. This event has already been examined using SDO data in \cite{Patsourakos2}. In what follows, we will present some new results. There are several important differences between this event and the previous one. Before initiation of the flare related to CME of 13 June 2010 and before the CME formation, we observe activation and a slow motion of a loop-like structure at the base of the future CME source region (Fig. 4(a1)). This is rather a loop of hot coronal matter than a filament, since it was not observed in H-alpha and HeI 1083 nm lines during the period preceding the CME initiation. In order to distinguish between such loops and eruptive filaments (prominences) formed by colder chromospheric matter, the term 'eruptive flux rope' was coined. It emerged that images of CME of 13 June 2010 in lines at 17.1 and 21.1 \textit{nm} differ significantly, as well as those of CME of 07 June 2011.  This is evident from brightness distribution in the region of the frontal structure, see Fig. 4(c4).  

Like in the previous event, several high-brightness loops are observed at the initial phase of the CME formation (e.g., Fig. 4 (a3, b3)). In this case, inner loops catch up with outer ones too, thus forming the CME frontal structure. Occurrence of the frontal structure is accompanied with occurrence of an extended dark area ('cavity').  
   
Fig. 5 (a, b) shows that, unlike in the previous event, velocity of the outer loop (or the CME frontal structure) reaches its maximum value within ~ 4 minutes in the SDO/AIA field of view (as has already been found in \cite{Patsourakos2}); subsequently, the CME velocity decreases fast (within ~ 7-9 minutes), too. Let us note one more difference between this event and the previous one: during the CME formation and its subsequent propagation, velocity of the eruptive flux rope is lower than that of the CME front in the greater part of its path.

Time profiles of velocity and acceleration of the CME front (as deduced from joint SDO/AIA, PROBA2/SWAP and SOHO/LASCO data) are presented in Fig. 5 (c, d). During this event, like during the previous one, instants of time when $V(t)$ and $I_{SXR}(t)$ as well as $(t)$ and $dI_{SXR}/dt$ reached their maximum values are close. There is a slight difference between duration of the periods when $V(t)$ and $I_{SXR}(t)$ reach their maximum values, too. 

No solid line V(t) in Fig. 5 (c) is due to no reliable data on V(t) between the last measured location of the CME front (SDO/AIA data) and the first measured location of the CME front (the LASCO field of view). Estimated average velocity in this interval of distances exceeds the maximum velocity value in Fig. 5 (a, c). This implies that there must be a hardly explicable maximum in this interval in V(t) profile. This problem can be solved if we assume, for instance, that CME propagates at a relatively large angle ($\approx$ $30^{\circ}-45^{\circ}$) to the plane of the sky at the initial stage of its propagation, and that the angle between the CME axis and the plane of the sky decreases. Consequently, comparison between the CME velocities (i.e., projection of 3-D velocity onto the plane of the sky) measured in the AIA and LASCO fields of view becomes incorrect. 

The qualitative features of formation and propagation of CMEs of 07 June 2011 and 13 June 2010 are typical of most events under consideration. The connection of their kinematic characteristics with soft X-ray intensity $I_{SXR}(t)$ from the flare region and with $dI_{SXR}/dt$ is typical, too. For the event of 27 March 2011, we did not manage to establish connection between $V(t)$ and $I_{SXR}(t)$, and $a(t)$ and $dI_{SXR}/dt$ because of the complex $I_{SXR}(t)$ profile.  Table 1 presents the main parameters describing kinematics of CMEs under consideration and its connection with soft X-ray emission from the flare region.
For CMEs of 08 March 2011 and 27 April 2011, decrease in velocity of the CME front V(t) after reaching its maximum value is relatively small (i.e., it is less than the arbitrary selected boundary of $100 km/s$). Let us exemplify it by the event of 08 March 2011, Fig. 6 (e). Velocity V(t) decreases by $\approx$ $95 km/s$ within 80 minutes, whereas the CME velocity on 07 June 2005, for instance, decreases by $\approx$ $570 km/s$ within 37 minutes (Fig.3(c)). Such variations in velocity with time influences the time profile of acceleration, see Fig. 6 (b): amplitude of absolute value of negative acceleration in this case is significantly lower than that of CMEs recorded on 13 June 2010 and 07 June 2011.

In \cite{Fain, FainZ}, limb CMEs \cite{Fain} and halo CMEs \cite{FainZ} with insignificant variation of velocity $V(t)$ after reaching the maximum value $V_{max}$ at moment $t(V_{max}$) form a special CME type. CMEs characterised by abrupt and dramatic variation $V(t)$ at $t>t(V_{max})$ form another CME type in these papers. CMEs, whose velocity after reaching the maximum value varies insignificantly, agree with the CME propagation model proposed in \cite{Zhang}.  According to findings in \cite{Temmer08, Temmer10, Patsourakos2, Fain, FainZ} and in this paper, there are many CMEs whose velocity decreases fast by more than 100 km/s after having reached the maximum.  The CME kinematics model proposed in \cite{Zhang} is not valid here. A more accurate model in this case is the model having 4 stages: there is also a stage of fast decrease in $V(t)$ between the stages of the main acceleration and the insignificant variation of velocity. 

Correspondence between initiation of the CME-related flare and initiation of motion of the eruptive filament (flux rope having another nature) in the region of the CME initiation is of importance when selecting a possible mechanism of the CME generation. Correspondence between the flare initiation and beginning of the main CME acceleration is of importance, too. Corresponding time is listed in Table 1.  We can draw the following conclusions: on 27 April 2011, motion of the eruptive filament or flux rope precedes the flare; on 13 June 2010, 11 February 2011, 27 March 2011 and 07 June 2011, it probably starts earlier. In these cases, determination accuracy of the initial moment of filament (flux rope) motion is insufficient. For the event of 08 March 2011, no eruptive filament or eruptive flux rope has been determined. As for the event of 27 April 2011, propagation of the CME front precedes the flare. It is possible that the standard flare model - CME \cite{Temmer10} - is valid for this event; according to this model, the CME propagation promotes initiation of a flare.

Let us now discuss some general results concerning the initial stage of the CME propagation. 

1) For all events under consideration, the average duration of the main CME acceleration $t_{acc}$ ~ 18.17 minutes was close to the average duration of enhancement of the soft X-ray intensity from the CME-associated flare region $t_{SXR}$ ~17.74 minutes (the event of 27 March 2011 was not taken into consideration in the second averaging). Here, $t_{acc}$ is the difference between the moment when CME velocity reaches its maximum $V_{max}$ and the moment when the first value of the CME velocity is determined from SDO data; $t_{SXR}$ is the duration of X-ray intensity enhancement $I_{SXR}(t)$, from the flare initiation to the maximum $I_{SXR}$ value. Besides, average values of the shortest duration of the main acceleration, combined with the shortest duration of soft X-ray enhancement (16.7 minutes), correspond to average values of the longest acceleration $V_{max}/t_{acc}$, combined with $V_{max}/t_{SXR}$ (0.98 $km/s^2$). Largest average values $t_{acc}$ combined with $t_{SXR}$ (27.6 minutes) correspond, conversely, to smallest average values $V_{max}/t_{acc}$ combined with $V_{max}/t_{SXR}$ (0.56 $km/s^2$). Both these results agree with results in \cite{Zhang}.

2). Dependences of the CME angular size $2 \alpha$ on time, as deduced from SDO and PROBA2 data, were plotted for 4 events (13 June 2010, 27 March 2011, 27 April 2011, 07 June 2011), see Fig. 7(a). Fig. 7(b) shows the CME angular size as an angle with vertex at the centre of the solar disc and with rays traced to the remotest points (in latitude) of the CME boundary. For all CMEs in Fig. 7(a), $2 \alpha$ increases with time. Range of variations in $2 \alpha$ is $\approx$ 1.5-3 for several ejections. The typical time of variation in the CME angular size (time of twofold variation in $2 \alpha$ since the first registration of CME) is $\approx$ 3.5-11 minutes.

3). Peculiarities of time variations in ratio of the CME height $d_H$ to its width $d_W$ have been revealed for the same 4 events (see item 2): $[d_H/d_W](t)$, Fig. 8(a). The CME height was counted from the CME source (arbitrarily considered as location of the CME-related flare, Fig. 8(b). During all the events presented in Fig. 8(a), $d_H/d_W$ decreased with time either since the first moment for which we managed to calculate ratio $d_H/d_W$ or since a later moment (see dependence for the event of 13 June 2010).  This implies that the CME broadening exceeds its extension in the longitudinal direction during this period. For the three events (except that of 07 June 2011), a tendency to an insignificant variation in $d_H/d_W$ with time is observed. This can be interpreted as the transfer to the self-similar expansion of CMEs when shape of the CME boundary is not changed with time.

\section{Discussion and conclusions} 
      \label{Discussion and conclusions} 
      Modern models of the CME formation are provided by a limited amount of experimental data on CMEs at early stages of their initiation and propagation \cite{Forbes}. The aim of this paper is to establish new regularities of the CME formation and of the initial stage of its propagation in order to use them for creating an adequate model of the CME initiation. This is possible due to data from instruments AIA aboard SDO that have unique characteristics, as well as due to joint data from PROBA2/SWAP, SOHO/LASCO and GOES.    
      
	We have established physical and morphological features of formation and initial stage of propagation of six CMEs: 
\begin{enumerate}
	\item event starts from the eruption of filament or loop-like structure having another nature.  This has been reliably established for five events of 13 June 2010, 11 February 2011, 27 March 2011, 27 April 2011, and 07 June 2011; 
	\item soon after the filament eruption several overlying loops occur and start moving at different velocities, thus forming the CME frontal structure (for all the events under consideration); 
	\item during the four events (13 June 2010, 11 February 2011, 27 March 2011, 07 June 2011), velocity of the CME front decreases abruptly by more than 100 km/s after having reached the maximum; it then varies insignificantly in the LASCO field of view. We suppose that such events form a special CME type. Velocity of two CMEs (08 March 2011 and 27 April 2011) decreases slightly after reaching the maximum value. Probably these CMEs represent another type of CMEs. According to the conclusion in \cite{Zhang}, this pattern of velocity variation is typical for CMEs. Findings in \cite{Temmer08, Temmer10, Patsourakos2, Fain, FainZ} and in this paper prove that the most typical is the profile of the CME velocity where velocity decreases abruptly by more than 100 km/s after having reached the maximum. 	
\end{enumerate}

The formed CME frontal structure is not the same in different spectral lines. The line $\lambda=17.1\;nm$ is formed at a lower temperature than the line $\lambda=21.1\;nm$. The shift of boundaries of the CME frontal structure along the radius implies that temperature of outer regions of the CME frontal structure is higher than that of inner regions.

	Variation in the CME angular size $ 2\alpha (t)$ demonstrates distribution and time variations in forces that influence the ejection boundary both outside and inside CME. The conclusion concerning the increase in the CME angular size $2\alpha$ between CME locations in the lower corona and in the LASCO field of view has been drawn in cite{Plunkett}. However, no data on typical temporal and spatial scales of such variations have been obtained. We managed to observe time variations in angular sizes of four CMEs in the SDO field of view. Value $2\alpha$ was found to increase up to threefold at the initial stage of propagation. Ratio of the CME angular size in the LASCO field of view to the first measured value $2\alpha$ could be $\approx$ 5 (for CME of 13 June 2010) and more. The typical temporal scale of doubling of the CME angular size after the first measurement of $2 \alpha$ was $\approx$ 3.5-11 minutes. In the frame of a simple CME model at the initial stage ('bubble', see \cite{Patsourakos2}), increase in the CME angular size with time evidences excess of internal force's influence on the CME boundary over external force's influence. 
	
	Study of ratio of the limb CME's longitudinal size to its transverse size $d_H/d_W$ provides solutions to several problems. First of all, this ratio is equivalent to the ratio of the CME velocity along its axis to the CME expansion velocity in the transverse direction. Knowledge of formation laws of this ratio allows us to estimate, for instance, propagation velocity of halo CMEs along their axes in three-dimensional space, by measuring the CME expansion velocity in the transverse direction (see discussion on this issue, based on LASCO data, in \cite{Michalek}). Besides, this ratio can be used to determine whether or not the CME expansion is self-similar. Comparison between CME propagation parameters and theory of self-similar expansion of CMEs \cite{Uralov} provides a more accurate estimate of self-similar CME propagation.  If ratio $d_H/d_W$ remains constant over time, this implies conservation of the CME form (self-similarity) over time. 
	
	Note that the parameter we have chosen to determine the CME expansion differs from that used in \cite{Patsourakos2} to perform the same analysis.  That parameter was referred to as 'aspect ratio' (ratio of the height of centre of circle $H_C$, which outlined well most of the CME boundary, to the radius of this circle $R_F$). Height $H_C$ was measured relative to the solar limb. We used the parameter closest to that in \cite{Michalek}. It can be shown that two these parameters should be connected by ratio $H_C/R_F \approx 2d_H/d_W-1$ for the event of 13 June 2010. When deriving this formula, we ignored deviation of location of the CME-related flare from the limb and from the centre of the CME basis.  Fig. 8 (c) compares dependence of $H_C/R_F$ on time \cite{Patsourakos2} with our dependence $[d_H/d_W](t)$ for the event of 13 June 2010. We see that $H_C/R_F$ variation over time is qualitatively close to dependence $[d_H/d_W](t)$. In the first approximation, relation between the two parameters is described by their above-stated ratio. Noteworthy is the fact that, starting from instant $t=$ 05:39:00, shape of the CME boundary does not correspond precisely to circle (e.g., compare the visible CME boundary at $t = $05:40:11 in our figure with circle in Fig. 3 from \cite{Patsourakos2}). 
	
	In conclusion, let us note that the use of data with high temporal and spatial resolution made it possible to obtain information about initiation and initial propagation stage of CMEs. However, a great deal needs to be done before initiation and development of CMEs become completely understood. In our next paper, we will try to relate the differences in $V(t)$ profiles established for 2 CME types to the peculiarities of the magnetic field and other features of active regions wherein CMEs initiate.

\begin{acks}
We are grateful to SDO/AIA, PROBA2/SWAP, SOHO/LASCO, \\
GOES and RHESSI teams for making their data available, and to V.V. Grechnev for fruitful discussions and his help in processing SDO data. This work was partially supported by grant No. 02.740.11.0576 under the Federal Programme 'Scientific and Scientific-Pedagogical Staff of Innovative Russia'.
\end{acks}

\mbox{}~\\

\bibliographystyle{spr-mp-sola}

\bibliography{paper}

\IfFileExists{\jobname.bbl}{} {\typeout{}
\typeout{****************************************************}
\typeout{****************************************************}
\typeout{** Please run "bibtex \jobname" to obtain} \typeout{**
the bibliography and then re-run LaTeX} \typeout{** twice to fix
the references !}
\typeout{****************************************************}
\typeout{****************************************************}
\typeout{}}

\end{article} 
\end{document}